\documentclass[preprint,12pt,sort&compress]{elsarticle}

\usepackage{amssymb,comment,float}
\usepackage[section]{placeins}
\usepackage[usenames,dvipsnames]{xcolor}
\usepackage{bm}

\usepackage{amsmath}

\journal{Acta Materalia}

\begin{document}

\begin{frontmatter}



\title{Automated Analysis to Reveal Grain Boundary Phase Microstructures} 

\author[1]{R. Daniel Moore}
\affiliation[1]{organization={Materials Physics, Sandia National Laboratories},
                city={Livermore},
                state={CA},
                country={USA}}
\ead{rdmoore@sandia.gov}
\author[1]{Ian S. Winter}
\author[2]{Robert E. Rudd}
\affiliation[2]{organization={Lawrence Livermore National Laboratory},
                city={Livermore},
                state={CA}, 
                country={USA}}
\author[3]{Fadi Abdeljawad}
\affiliation[3]{organization={Department of Materials Science and Engineering, Northwestern University},
            city={Evanston},
            state={IL},
            country={USA}}
\author[2]{Timofey Frolov}
\ead{frolov2@llnl.gov}

\begin{abstract}
We develop a method for analyzing grain boundary (GB) microstructures that identifies distinct interfacial phases and the dislocation line defects separating them. Similar to bulk materials, GBs can adopt multiple distinct interfacial phases and undergo first-order phase transitions. When these phases coexist, their spatial arrangement and phase junctions constitute a GB microstructure, characterized by variations in excess properties and line defects with associated dislocation content. Despite this intrinsic heterogeneity, our ability to quantitatively characterize GB microstructures remains limited, as it requires identification of individual GB phases, phase-resolved excess properties, and the Burgers content of phase junctions — capabilities not available in existing automated methods. Here, we present an automated tool that performs interfacial microstructure mapping for planar coincidence site lattice GBs. The method identifies the spatial distribution of GB phases, quantifies phase-specific excess properties, and estimates the Burgers content of GB phase junctions. We demonstrate the approach using three representative cases: (i) quantification of mass transport during diffusion-limited GB phase transformations; (ii) identification of structurally indistinguishable phases formed by vacancy and interstitial loops; and (iii) characterization of GB microstructures containing phase nuclei. More broadly, this framework enables quantitative studies of GB evolution processes, including spinodal decomposition and coarsening with direct implications for GB deformation, creep, and migration.
\end{abstract}



\begin{keyword}
Grain Boundaries \sep Phase Junctions \sep Microstructure


\end{keyword}

\end{frontmatter}



\section{Introduction}
Grain boundaries (GBs) are well known for their influence on material properties and performance~\cite{hirth1972influence,mishin1997grain,watanabe1994impact,watanabe1983grain,humphreys2012recrystallization}. Analogous to bulk phases, GBs can exist in multiple distinct structural states, referred to as GB phases~\cite{frolov2026thermodynamics}, each characterized by unique excess thermodynamic quantities. Transitions between these states may occur via first-order interfacial phase transformations~\cite{frolov2013structural,winter2022nucleation}, leading to discontinuous changes in the corresponding excess properties~\cite{dillon2008relating}. Distinct GB phases are separated by one-dimensional interfacial defects known as GB phase junctions (GBPJs) that possess dislocation content~\cite{frolov2021dislocation,hirth1996steps,dillon2026basic}.

Over the past decade, compelling evidence for GB phase coexistance, and phase transitions have been observed across a wide range of materials in both computational and experimental studies~\cite{frolov2013structural,choi2025faceting,zhu2018predicting,frolov2015segregation,chen2025quasiaperiodic,frolov2018grain,frolov2018structures,chen2024grand,hooshmand2021twin,zhou2025boron,devulapalli2024topological,ding2025hierarchy,langenohl2022dual,GBPhaseExp,peter2018segregation,frolov2026thermodynamics}. Atomistic simulations have been used to explore GB phases and phase transitions in FCC~\cite{freitas2018free,frolov2013structural,choi2025faceting,zhu2018predicting,frolov2015segregation}, BCC~\cite{chen2025quasiaperiodic,frolov2018grain,frolov2018structures}, and HCP~\cite{chen2024grand,hooshmand2021twin} materials, while direct observation of first-order structural transformations has been achieved through high-resolution transmission electron microscopy in a range of elemental and alloyed systems~\cite{zhou2025boron,devulapalli2024topological,ding2025hierarchy,langenohl2022dual,GBPhaseExp,peter2018segregation}. Beyond these observations of GB phase transitions, experimental studies of deformation-modified GBs have revealed substantial changes in diffusion kinetics, solute segregation, and microstructural stability, suggesting that local structural and chemical heterogeneity within GBs may play an important role in determining their properties~\cite{glienke2020grain,sankaran2020grain,divinski2015effect,wilde2023internal}.

Nonequilibrium GB processes, such as migration, sliding, and defect absorption, are expected to be mediated by nucleation and motion of disconnections~\cite{chen2020temperature,khater2012disconnection,thomas2026grain,gautier2025quantifying}. However, it has recently been demonstrated that GBPJ's have Burgers vectors smaller in magnitude than those of regular GB disconnections~\cite{winter2025quantifying}, making GBPJs potentially more energetically favorable. As a result, processes typically attributed to disconnections may instead proceed via GBPJ formation, leading to the emergence of patches of distinct GB phases separated by networks of these line defects~\cite{wei2021direct}. This behavior is analogous to partial dislocation formation in bulk crystals, where the total energy is lowered through dissociation into segments with smaller Burgers vectors. Thus, it should be expected that under non-equilibrium processing conditions, GBs can develop complex interfacial microstructures composed of distinct GB phases separated by networks of GBPJs. 

Quantitatively characterizing such interfacial microstructures remains challenging, particularly in extracting GB phase excess properties while simultaneously resolving the spatial distribution, and Burgers vector content, of individual GBPJs. Although previous studies have developed tools to compute GB excess properties for bicrystals containing uniform GB phases across large datasets using grand-canonical sampling~\cite{chen2024grand,zhu2018predicting}, extending these approaches to mapping and tracking GB phase evolution within a microstructure containing different phases and GBPJs remains a significant challenge and is reliant on manual analysis by Burgers circuit construction~\cite{GBPhaseExp,winter2024phase,frolov2018grain}. 

In recent years, significant effort has been devoted to extracting interfacial line defect information from atomistic data, leading to the development of computational approaches such as extensions of the Dislocation Extraction Algorithm (DXA)~\cite{stukowski2012automated} and methods based on evaluation of the Nye tensor~\cite{hartley2005representation,winter2022characterization,Winter2026}. DXA localizes dislocation lines using Burgers circuits and determines their Burgers vectors, providing an unambiguous description of lattice dislocations. Although it has been extended to interfacial systems, its application remains limited because it requires explicit knowledge of the relaxed interfacial reference structure, which is often unavailable. Nye-tensor-based approach circumvents this requirement but depends on defining the Displacement Shift Complete (DSC) lattice~\cite{bollmann2012crystal} and does not provide the precise line position, or disconnection step height. More recently, the Interfacial Line Defect Analysis (ILDA) method~\cite{deka2023automated} introduced an automated procedure for extracting interfacial dislocations and disconnections by constructing Burgers circuits on a surface mesh of coincidence-site (CIS) atoms. While ILDA enables automated measurement of both Burgers vectors and step heights, it cannot detect line defects in more general GBs with complex structures, such as those considered in this work, where CIS atoms cannot be reliably identified and, as a result, the Burgers-circuit mesh cannot be constructed.

Despite substantial progress in tools for interfacial phase and line defect analysis, existing methods remain limited in their ability to robustly and systematically quantify GB microstructures containing multiple phases and phase junctions. In this work, we introduce an automated framework that (i) identifies distinct GB phases, (ii) computes their excess thermodynamic properties, and (iii) detects and characterizes interfacial line defects in planar coincidence-site lattice (CSL) boundaries. We demonstrate the approach on both tilt and twist bicrystals in copper and tungsten. Finally, we discuss the relevance of the presented examples to GB mediated non-equilibrium processes including migration and creep.

\section{Methodology}
\label{sec:Methodology}

We begin by introducing the key quantities used to characterize GB phases and the line defects between them. We consider a planar GB (lying in the x-y plane) that contains coexisting GB phases separated by interfacial line defects, such as GBPJs and disconnections. Spatially resolved interfacial excess thermodynamic properties are used to identify and characterize distinct GB phases, while the Burgers vector content and line direction are used to characterize the line defects separating them.


\subsection{Interfacial Excess Quantities}
\begin{figure}[H]
    \centering
    \includegraphics[width=0.3\textwidth]{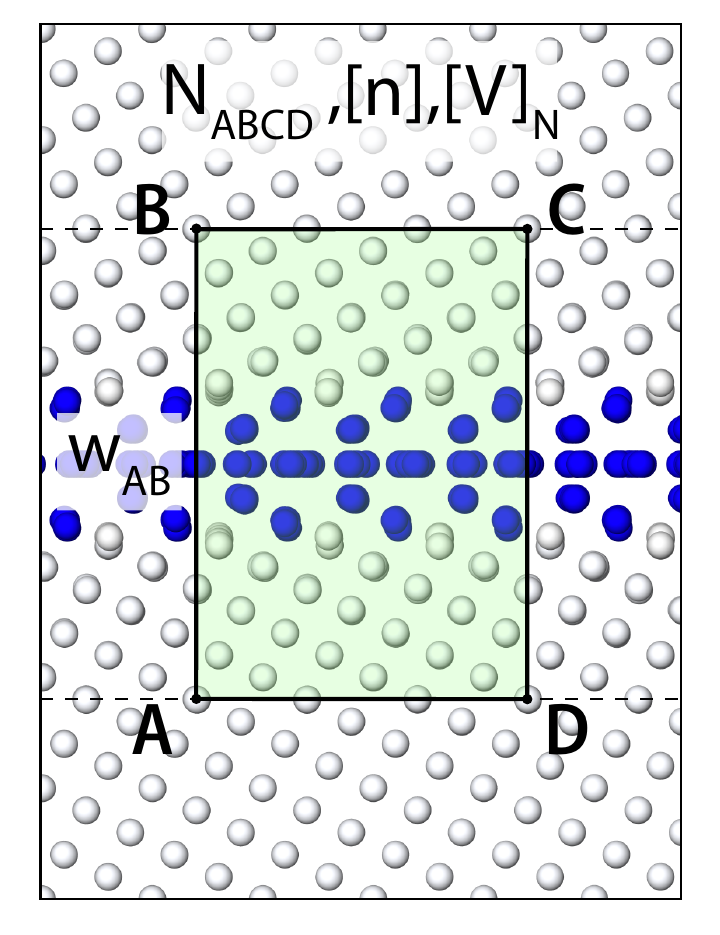}
    \caption{An example of a rectangular analysis region (ABCD) enclosing the grain boundary and adjacent crystal regions is translated along the interface. Within each window, the number of atoms $N_{ABCD}$, excess number of atoms $[n]$, excess volume $[V]_N$, and crossing vector $\mathbf{w}_{AB}$ are evaluated to characterize the local grain-boundary structure.}
    \label{FIG:0}
\end{figure}

GB phases are characterized by a distinct set of thermodynamic excess quantities, including the excess volume $[V]_N$, excess shears $[\bm{B}]_N$ associated with the relative translation of the adjoining grains, and the number of atoms $[n]$~\cite{frolov2012thermodynamics}. To resolve spatial variations in these quantities within heterogeneous GB microstructures, we define a parallelepiped region with corners attached to bulk atoms corresponding to the reference lattice enclosing the boundary and portions of the adjoining grains, and translate it along the interface plane. Within each window, the excess quantities are evaluated and then smoothed with a moving average approach, yielding local profiles that resolve spatial variations in excess properties and, thus, GB structure. The analysis region and the quantities evaluated within it are illustrated in Fig.~\ref{FIG:0}.

The excess volume contributes to the relative grain displacement normal to the boundary plane and is calculated as~\cite{cahn1979thermodynamics}:
\begin{equation}
    [V]_N = \frac{V_{\mathrm{GB}} - N_{\mathrm{GB}}\,\Omega}{A_{\mathrm{GB}}},
    \label{eq:excess_volume}
\end{equation}
where $V_{\mathrm{GB}}$ is the total Voronoi volume of all atoms within the analysis region, $N_{\mathrm{GB}}$ is the number of atoms in the region, $\Omega$ is the bulk atomic volume, and $A_{\mathrm{GB}}$ is the interfacial area contained within the analysis region.

The second GB property that also contributes to a normal displacement of one grain relative to the other is the number of atoms, $[n]$, originally introduced for symmetric tilt and twist GBs as~\cite{frolov2013structural} 
\begin{equation}
[n]=\frac{N_{\mathrm{total}}\ \mathrm{mod}\ N_{\mathrm{plane}}^{\mathrm{bulk}}}
{N_{\mathrm{plane}}^{\mathrm{bulk}}}\in[0,1),
\label{eq:n_Cal}
\end{equation}
where $N_{\mathrm{total}}$ is the total number of atoms in the bicrystal and $N_{\mathrm{plane}}^{\mathrm{bulk}}$ is the number of atoms in a bulk atomic plane parallel to the grain boundary. A general definition of $[n]$ that includes asymmetric GBs is given in Ref.~\cite{winter2025quantifying}.  Computing $[n]$ using Eq.~\ref{eq:n_Cal} is straightforward for simulation cells with periodic boundary conditions along the GB, but it is not well suited for a moving local analysis region. As this analysis volume translates through the bicrystal, atoms continuously enter and leave the region, introducing artificial fluctuations in the atom count and, consequently, in the computed excess number of atoms. Rather than relying on direct atom counting, we obtain $[n]$ from non-DSC components of crossing vectors~\cite{winter2025quantifying}. The relative translation between the two grains is uniquely described by the Wigner--Seitz translation vector, $\mathbf{t}^{\mathrm{WS}}$. It is obtained by mapping any crossing vector, $\mathbf{w}$, connecting two lattice sites in the adjoining grains into the Wigner--Seitz cell of the DSC lattice. This mapping removes the ambiguity associated with adding DSC lattice vectors, yielding a unique representation of the relative grain disregistry. Any crossing vector can be decomposed as~\cite{winter2025quantifying}
\begin{equation}
\mathbf{w}=\mathbf{t}^{\mathrm{WS}}+\mathbf{d}^{\mathrm{DSC}},
\label{eq:w_decomp}
\end{equation}
where $\mathbf{d}^{\mathrm{DSC}}$ is the DSC lattice vector nearest to $\mathbf{w}$, ensuring that $\mathbf{t}^{\mathrm{WS}}$ resides within the Wigner--Seitz cell of the DSC lattice~\cite{winter2025quantifying}. In the present analysis, $\mathbf{w}$ is obtained by selecting pairs atoms, which correspond to lattice sites in the reference configuration, from planes above and below the grain boundary at the two ends of the analysis region, such as vector $\mathbf{w}_{AB}$ shown in Fig.~\ref{FIG:0}. The displacement vectors connecting these atom pairs provide independent estimates of the crossing vector, and the average over ten pairs is used to improve accuracy.

The excess number of atoms associated with the GB can then be determined directly from the normal component of the Wigner--Seitz translation vector and the excess volume~\cite{winter2025quantifying}:
\begin{equation}
[n]=\frac{t^{\mathrm{WS}}_3-[V]_N}{d_3^{\mathrm{DSC}}},
\label{eq:n_geometric}
\end{equation}
where $t^{\mathrm{WS}}_3$ is the component of the Wigner--Seitz translation vector normal to the GB and $d_3^{\mathrm{DSC}}$ is the bulk interplanar spacing normal to the interface related to the number of atoms per plane per unit area.

Finally, we calculate the \emph{relative difference in the number of atoms}, $\Delta N$, which quantifies the absolute change in the number of atoms contained in the analysis region as it moves from one GB region to another. Following Ref.~\cite{frolov2021dislocation}, it is defined as
\begin{equation}
    \Delta N = \frac{N_{\mathrm{i}} - N_{\mathrm{ref}}}{N_{\mathrm{bulk}}},
\end{equation}
where $N_{\mathrm{i}}$ and $N_{\mathrm{ref}}$ are the numbers of atoms within the bounding boxes enclosing the GB phase of interest and a chosen reference GB phase, respectively, and $N_{\mathrm{bulk}}$ is the number of atoms in an equivalent bulk atomic plane parallel to the GB.

\subsection{Identification of GB Phase Junctions}
\label{sec:pj}
When two GB phases meet at a phase junction, differences in their equilibrium atomic arrangements generate a localized strain field that accommodates the structural mismatch between the adjoining phases. This mismatch alters the relative spacing of atoms within the bulk planes adjacent to the GB. Consequently, line defects within the GB plane can be located by tracking spatial variations in the local atomic separation along these planes.

\begin{figure*}[!ht]
	\centering
	\includegraphics[width=1\textwidth]{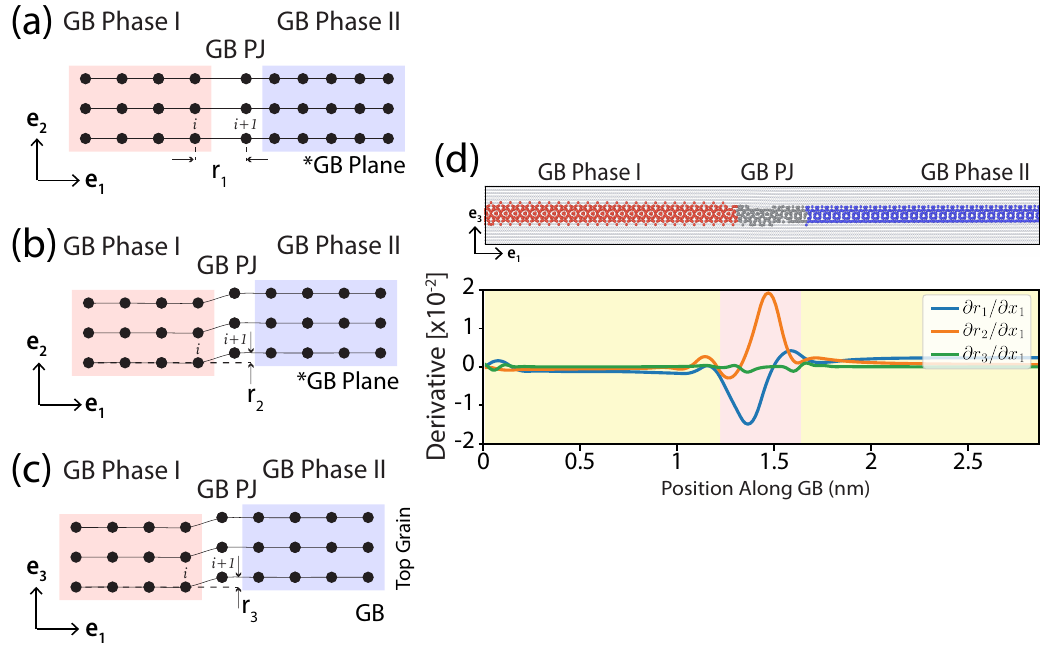}
	\caption{Schematic illustration of the displacement induced in the surrounding bulk lattice planes adjacent to a grain-boundary phase junction (GBPJ). (a–c) Examples illustrating the relative lattice displacements (and the separation vector component) associated with a general GBPJ. (d) Representative periodic GB containing two distinct GB phases (red and blue) separated by a GBPJ (top), together with the corresponding spatial derivatives of the separation-vector components (bottom). Yellow-shaded regions indicate the GB phases, while the pink-shaded region denotes the GBPJ.}
	\label{FIG:PJPosition}
\end{figure*}

Figure~\ref{FIG:PJPosition}(a--c) schematically illustrates how a GBPJ perturbs the positions of atoms within bulk planes adjacent to the GB. For consecutive atoms along a selected plane parallel to the boundary, the local separation vector is evaluated while moving across the junction. Within a uniform GB phase, the components of this vector remain approximately constant. Crossing a GBPJ, however, changes the relative arrangement of the neighboring atoms, producing a localized variation in the atomic separation. The corresponding changes in the $\textbf{e}_1$, $\textbf{e}_2$, and $\textbf{e}_3$ components are indicated by the arrows in Fig.~\ref{FIG:PJPosition}(a--c), respectively.

In practice, the GB region is first located by spatially averaging the coordinates along $\textbf{e}_3$ of atoms classified as non-FCC or non-BCC using common neighbor analysis~\cite{honeycutt1987molecular}. These atoms are then excluded, leaving only the crystalline atoms in the adjoining grains. The bulk atomic planes adjacent to the GB are enumerated, and the plane nearest the boundary is selected for analysis. For each pair of consecutive atoms (lattice sites) along this plane, the components of the separation vector, $\mathbf{r}^{(i)}$, are calculated directly along the basis directions $\left(\mathbf{e}_1,\mathbf{e}_2,\mathbf{e}_3\right)$ as
\begin{equation}
    r_{k}^{(i)} = x_{k}^{(i+1)}-x_{k}^{(i)},
    \qquad k=1,2,3,
    \label{eq:plane_separation}
\end{equation}
where $x_{e_k}^{(i)}$ denotes the coordinate of atom $i$ along the cartesian directions in the current configuration.

Within a uniform GB phase, the components of $\mathbf{r}$ vary smoothly and remain approximately constant. In contrast, at a GBPJ, the relative arrangement of neighboring atoms changes, producing a localized variation in the separation vector. Consequently, the spatial derivatives of its components exhibit peaks near the junction location.

To suppress atomistic noise, the discrete separation-vector components $r_{k}^{(i)}$ are grouped into bins with a width of 1~nm, and the mean value within each bin is computed. A cubic spline is then fitted through the bin centers to obtain the continuous fields $r_{k}(x_j)$, where $x_j$ denotes the spatial coordinate along an in-plane cartesian direction $\mathbf{e}_j$. The spatial derivatives of these fields are subsequently evaluated as
\begin{equation}
    \frac{\partial r_{k}}{\partial x_j},
    \qquad k=1,2,3,
\end{equation}
where $k$ denotes the separation-vector component along $\mathbf{e}_k$. GB phases are identified as connected intervals along the boundary in which all three derivatives remain within a threshold $\pm\tau$, where $\tau = 4\,\mathrm{MAD}$. MAD represents the median absolute deviation and is calculated over all separation vector derivative values as~\cite{nist_mad}
\begin{equation}
    \mathrm{MAD} = \operatorname{median} \left| g-\tilde{g} \right|  ,
\end{equation}
where $g$ represents $\partial r_{k}/\partial x_j$. The complementary regions are classified as GBPJs, while the adjoining GB phases are retained for the subsequent Burgers-circuit analysis. Figure~\ref{FIG:PJPosition}(d) demonstrates this procedure for a representative GBPJ in W, where sharp peaks in the separation-vector derivatives identify the phase-junction locations. The detected GB phases and GBPJs are shown in yellow and pink, respectively.

\FloatBarrier

\subsection{Burgers Vector Analysis}
\begin{figure*}[th!]
	\centering
	\includegraphics[width=1\textwidth]{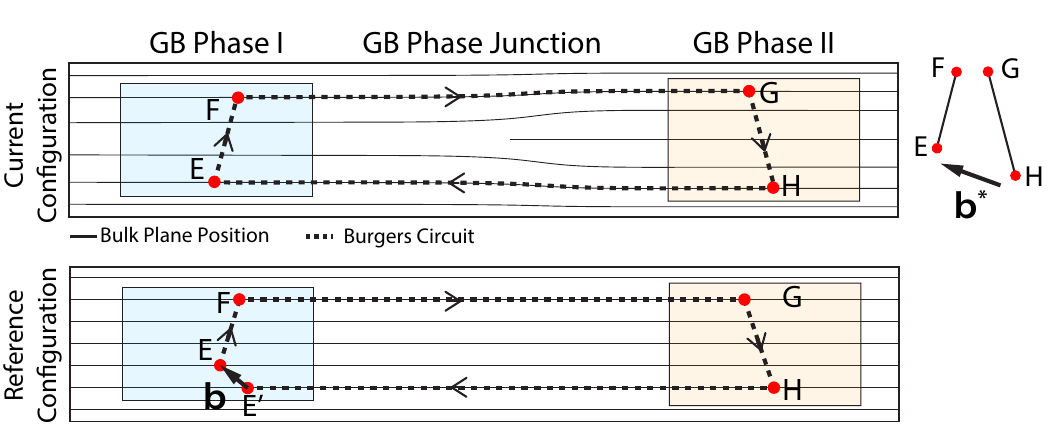}
	\caption{(Top) Schematic of a circuit construction around a general grain boundary phase junction in the current configuration. The approximated Burgers vector, $\textbf{b}^*$ is computed as sum of the crossing vectors. (Bottom) The same circuit mapped to the reference configuration with the illustrated closure failure.}
	\label{FIG:circuit}
\end{figure*}
Once the location of a GBPJ and the GB phases on either side has been identified, we \textit{approximate} the Burgers content of each junction by constructing closed circuits in the current configuration. For the proceeding analysis we adopt the finish-to-start right-hand convention~\cite{anderson2017theory}, all of our circuits are drawn in a clockwise fashion. For the closed circuit, presented in Fig.~\ref{FIG:circuit}, the line integral along the contour EFGH in the current configuration satisfies 
\begin{equation}
    \mathbf{0} =\oint_c d\mathbf{x} =\mathbf{w}^{EF} +\mathbf{w}^{FG}+\mathbf{w}^{GH}+\mathbf{w}^{HE},
\end{equation}
where \textbf{x} denotes the positions in the \textit{current} configuration and $\textbf{w}^{AB}=\textbf{x}^B-\textbf{x}^A$ is the vector associated with a segment, \textbf{AB}, along the contour. 

In a conventional Burgers-circuit construction, the corresponding circuit is mapped on to a reference configuration where the resulting closure failure defines the Burgers vector,
\begin{equation}
    \textbf{b} =\oint_C d\textbf{X} =\textbf{W}^{EF} +\textbf{W}^{FG}+\textbf{W}^{GH}+\textbf{W}^{HE'},
\end{equation}
where \textbf{X} represents the positions in the reference configuration and $\textbf{W}^{AB}=\textbf{X}^B-\textbf{X}^A$ is the vector associated with segment \textbf{AB} in the reference configuration. We choose our circuit such that $\bm{W}^{FG}=-\bm{W}^{HE'}$.

Because the reference configuration is not always available during an on-the-fly analysis, we make the assumption that the crossing vectors in the reference and current configurations are approximately equivalent, 
\begin{equation}
\textbf{W}^{EF} \approx \textbf{w}^{EF}, \qquad \textbf{W}^{GH} \approx \textbf{w}^{GH},
\end{equation}
Under this assumption, we can then define an \textit{approximated} Burgers vector as 
\begin{equation}
\textbf{b}^* = \left(\textbf{w}^{EF}+\textbf{w}^{GH}\right),
\label{eq:approxBurgers}
\end{equation}
Accordingly, $\textbf{b}^*\approx\textbf{b}$ to the extent that the measured current configuration crossing vectors provide an accurate approximation to their reference configuration counterparts. 

To construct the circuit shown in Fig.~\ref{FIG:circuit}, four corner atoms,
$\mathbf{E}$, $\mathbf{F}$, $\mathbf{G}$, and $\mathbf{H}$, are selected from bulk reference planes located $n$ lattice spacings above and below the GB. Starting from $\mathbf{E}$ in the lower grain within phase~I, the circuit crosses the GB to $\mathbf{F}$ in the upper grain, proceeds along the upper bulk plane across the GBPJ to $\mathbf{G}$ within phase~II, crosses the GB to $\mathbf{H}$ in the lower grain, and returns along the lower bulk plane to $\mathbf{E}$, closing the circuit.

The four corner atoms are selected as follows:
\begin{enumerate}
    \item $\mathbf{E}$ is selected at random\footnote{A random number generator is used to select an atom ID from a list of available atoms for \textbf{E}} from atoms on the lower reference
    plane within a 1~nm-wide region centered in phase~I.
    \item $\mathbf{F}$ is selected as the atom (corresponding to a lattice site) in the upper grain lying closest to vertical above $\mathbf{E}$ and nearest to the upper reference plane. 
    \item $\mathbf{G}$ is chosen at random within a 1~nm-wide region at the center of phase~II on the same upper reference plane as $\mathbf{F}$ and on the row (along the $\textbf{e}_2$ direction) as $\mathbf{F}$. The integer number of lattice steps $n_{\mathrm{steps}}$ along the plane from $\mathbf{F}$ to $\mathbf{G}$ is then determined by counting the nearest-neighbor gaps in the direction of $\mathbf{F} \to \mathbf{G}$. 
    \item $\mathbf{H}$ is found by advancing along the lower reference plane from $\mathbf{E}$ in the $+\textbf{e}_1$ direction by exactly $n_{\mathrm{steps}}$ lattice intervals. 
\end{enumerate}

Using our prior assumption that the crossing vectors in the current configuration, namely $\textbf{w}_{FE}$ and $\textbf{w}_{HG}$, are approximately equivalent to their counterparts in the reference configuration, the Burgers vector can be estimated from Eq.~\ref{eq:approxBurgers}.  
The corner-selection procedure is repeated for $N_c$ independent circuits, each with new random choices of $\mathbf{E}$ and $\mathbf{G}$. Circuits returning $|b_i| > \tfrac{3}{4} a$, where $a$ is the bulk lattice parameter, indicating that the circuit has either failed to fully enclose the defect or that the geometric closure between $\mathbf{F} \to \mathbf{G}$ and $\mathbf{E} \to \mathbf{H}$ has broken down, are discarded and replaced by new draws. The mean and standard deviation of the surviving circuits define the reported Burgers vector and its uncertainty,
\begin{equation}
    \langle \mathbf{b} \rangle = \frac{1}{N_c} \sum_{i=1}^{N_c} \mathbf{b}_i, \qquad
    \sigma_{\mathbf{b}} = \mathrm{std}\!\left( \{ \mathbf{b}_i \} \right).
\end{equation}
We reiterate that the Burgers vector computed by the present method is an approximate measure of the relative displacement in the lattice associated with a grain-boundary phase junction. While the construction uses information only from the current configuration, and is therefore not a rigorous topological Burgers vector, it provides a robust and fully automated characterization of GBPJ line defects that is currently unavailable with existing approaches.

\section{Results}
\label{sec:Results}
\subsection{Example of GB Phase Coexistence in a $\Sigma5$ Tilt Boundary}

\begin{figure*}[h!]
	\centering
	\includegraphics[width=1\textwidth]{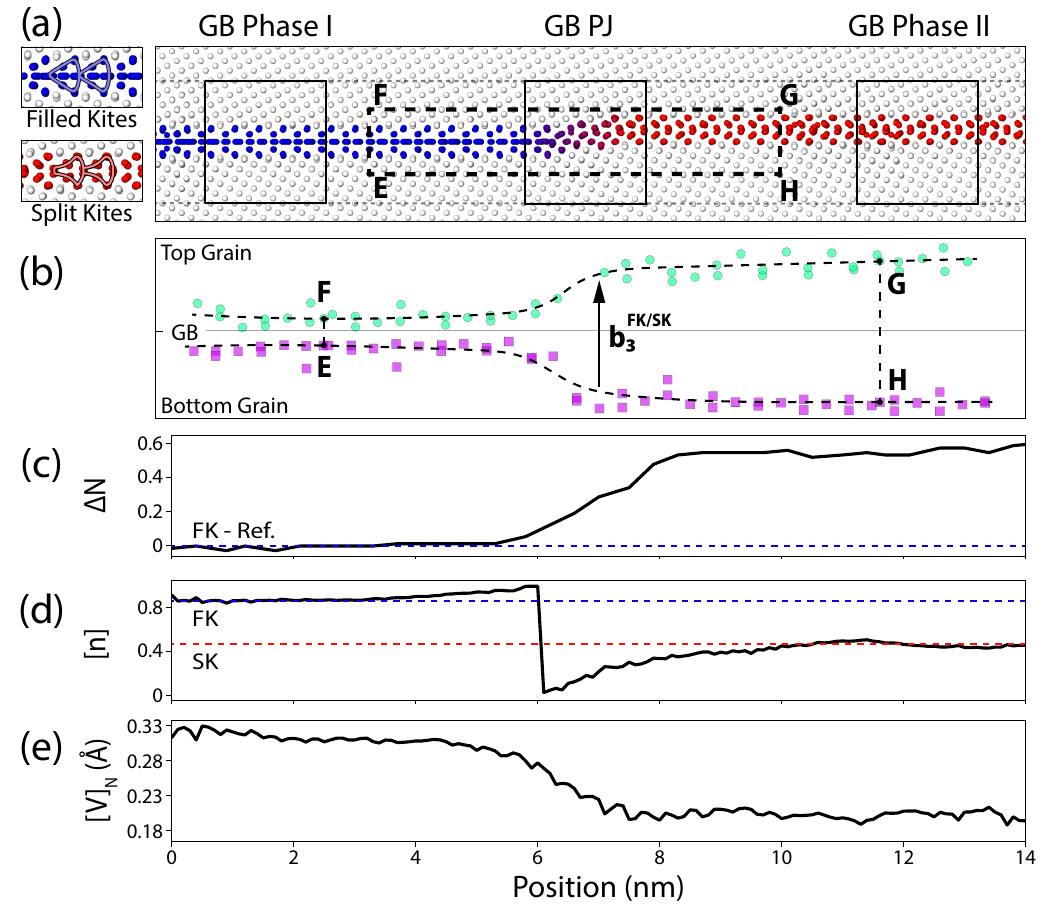}
	\caption{Automated analysis of GB phases, their excess thermodynamic properties, and the dislocation content of the GB phase junction, demonstrated for a heterogeneous Cu $\Sigma5$(310)[001] tilt boundary containing filled-kite (FK, blue) and split-kite (SK, red) phases. (a) Schematics of the moving analysis window and a trial Burgers circuit E-F-G-H are illustrated by the solid and dashed lines, respectively. (b) Magnified out-of-plane displacements of the bulk planes adjacent to the GB in the top and bottom grains, along with the constructed Burgers circuit and resultant Burgers vector $\textbf{b}^{FK/SK}$. (c--e) Profiles of the relative difference in the number of atoms $\Delta N$, GB number of atoms $[n]$, and excess volume $[V]_N$, respectively.}

	\label{FIG:1}
\end{figure*}
We first demonstrate the method on a well-characterized model system: a heterogeneous Cu $\Sigma5$(310)[001] symmetric tilt boundary containing two coexisting GB phases, the filled-kite (FK) and split-kite (SK) structures, shown in Fig.~\ref{FIG:1}(a). This boundary provides an ideal test case because its phases, their excess properties, and the Burgers content of the junction separating them have been characterized by prior manual analysis~\cite{frolov2021dislocation}. The FK and SK phases have different atomic densities, $[n] = 6/7$ and $[n] = 7/15$ respectively, meaning that the SK phase requires insertion of $\approx 3/5$ fraction of a bulk atomic plane relative to the FK phase. When these phases coexist on the same boundary plane, they are separated by a GBPJ whose Burgers content is normal to the GB plane.

The dislocation content of the junction is revealed by the out-of-plane displacements of the bulk atomic planes adjacent to the GB, shown in Fig.~\ref{FIG:1}(b). Because the SK phase accommodates a higher density of atoms than the FK phase, the planes bounding the SK region are displaced 
farther apart than those bounding the FK region. The Burgers circuit constructed around the junction  yields a purely normal Burgers vector component $b_3 = 0.35~\text{\AA}$, consistent with prior results obtained by manual circuit construction for this boundary~\cite{frolov2021dislocation}. A sensitivity analysis of $b_3$ as a function of distance from the GBPJ is provided in the supplemental material. The non-zero normal component indicates that GBPJ motion during a FK-to-SK transformation must proceed via climb, requiring a supply of atoms a topic explored in later sections.

Figs.~\ref{FIG:1}(c--e) show the spatially resolved excess quantities obtained from the moving window analysis. The difference in the number of atoms $\Delta N$, using the FK phase as reference, is shown in Fig.~\ref{FIG:1}(c). The profile is flat within the FK region and rises sharply across the junction to a plateau of approximately $0.6$ of a bulk atomic plane within the SK region, confirming that the SK structure was created from the FK phase by absorbing extra atoms. The GB atomic excess $[n]$, shown in Fig.~\ref{FIG:1}(d), takes values of $6/7$ (or equivalently $-1/7$) and $7/15$ in the FK and SK regions respectively, in agreement with the expected ground-state values for these phases. The excess volume profile, Fig.~\ref{FIG:1}(e), yields corresponding values of $\approx 0.31~\text{\AA}$ in the FK region and $\approx 0.20~\text{\AA}$ in the SK region, reflecting the different atomic arrangements of the two phases, which is consistent with previously reported values~\cite{frolov2021dislocation}. Together, the profiles in Figs.~\ref{FIG:1}(c--e) demonstrate that the method simultaneously resolves the spatial distribution of coexisting GB phases, their phase-specific excess thermodynamic properties, and the dislocation content of the junction separating them.

Having validated the method on this model system, we proceed by demonstrating its application to other GB phenomena that involve GB phase transformations. We first track mass transport during a diffusion-limited GB phase transformation. We then demonstrate the ability to distinguish different GBPJs formed by two structurally identical GB phases. Finally, we extend the analysis to more general cases: a GBPJ with both normal and shear Burgers components, and a two-dimensional GB microstructure containing a GB phase nucleus.

\subsection{Kinetics and Mass Transport of a Diffusion-Limited GB Phase Transformation}
\begin{figure*}[h!]
    \centering
    \includegraphics[width=1.0\textwidth]{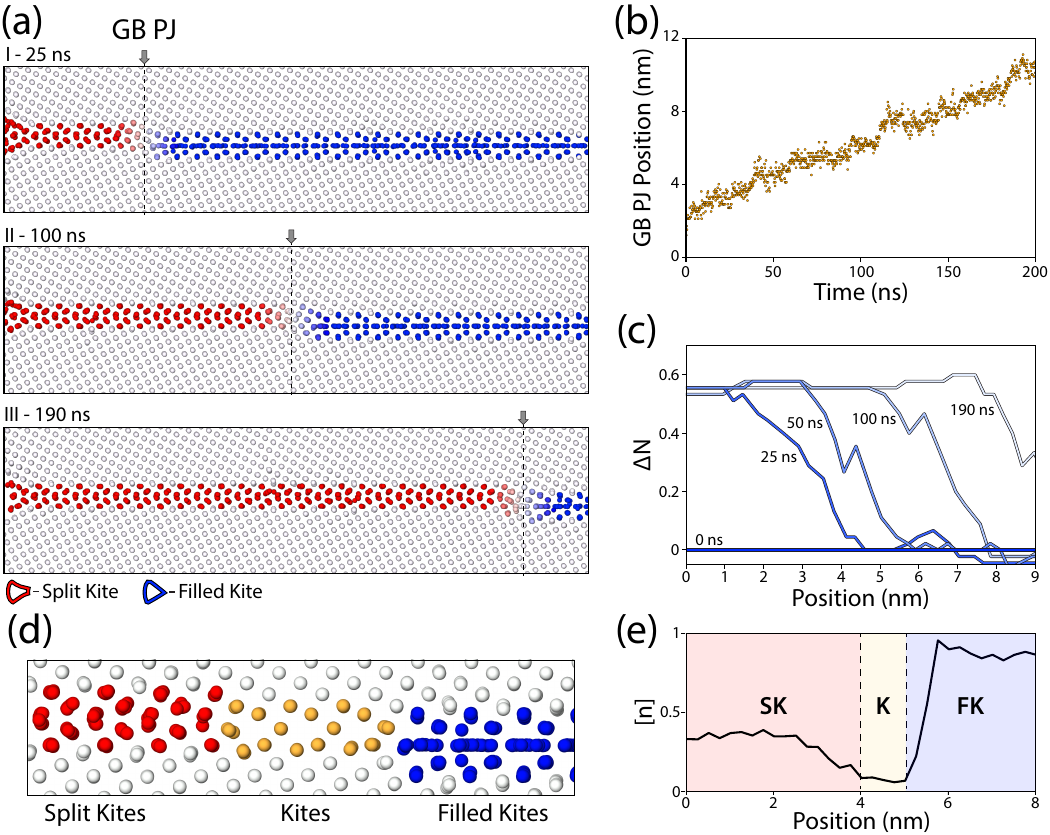}
    \caption{Analysis of the diffusion-limited GB phase transformation. (a) Atomic snapshots of the FK-to-SK GB phase transition in a simulation with free surface at 25, 100, and 190 ns. (b) GBPJ position as a function of time. (c) Profiles of the change in the number of GB atoms during the transformation, illustrating an insertion of extra atoms from the free surface equivalent to 3/5 of an atomic plane, accommodated by the GB  through a first-order GB phase transformation. (d) Transient formation of a regular kite phase at the transforming GB front. (e) Profile of GB atomic excess across a junction containing the intermediate kite phase.}
    \label{FIG:3}
\end{figure*}
We apply our method to quantify the kinetics of a first-order, diffusion-limited GB phase transformation in the same Cu $\Sigma5$(310)[001] tilt boundary modeled using an embedded atom method (EAM) potential~\cite{mishin2001structural}. The boundary is initially constructed with a FK phase ($[n] = -1/7$), after which a free surface is introduced at the left edge of the simulation cell to allow for atomic diffusion into the GB plane. The system is subsequently annealed in the canonical ensemble at 800~K for 200~ns using a 2~fs timestep (a video is provided in the supplemental material). To visualize the atomic structure, intermediate configurations are extracted during annealing, rescaled to the 0~K lattice parameter, and energy-minimized. Fig.~\ref{FIG:3}(a) shows snapshots at 25, 100, and 190~ns during the anneal, revealing the first-order structural transformation from FK to SK, which proceeds by nucleation and growth of the SK phase at the surface triple junction. The SK phase has a higher atomic density of $[n] = 7/15$, requiring $\approx 3/5$ of a bulk atomic plane to be supplied by the surface. 

The analysis identifies the GBPJ position from snapshots extracted at 0.2~ns intervals, plotted versus annealing time in Fig.~\ref{FIG:3}(b). The resulting trajectory exhibits fluctuations reflecting the underlying atomistic mechanism: because the two GB phases have different atomic 
densities, atoms must be supplied from the free surface, diffuse along the GB, and be delivered to the GBPJ before the junction can advance. GBPJ motion is therefore coupled to the intermittent supply and absorption of atoms, producing transient fluctuations in its position. We note that the present approach enables, in principle, calculation of GBPJ mobility when the driving force is known; this will be the subject of future work.

To quantify this mass transport, we track the change in the number of atoms, $\Delta N$, relative to the FK phase for snapshots at 25, 50, 100, and 190~ns, as shown in Fig.~\ref{FIG:3}(c). The analysis confirms that the expected fraction of $\approx 0.6$ of a bulk atomic plane is incorporated into the boundary as the transformation to the SK phase proceeds.

Finally, we observe that a GBPJ can dissociate into two phase junctions, analogous to the dissociation of a perfect dislocation into two partial dislocations. Unlike conventional dislocation dissociation, however, this process is mediated by mass transport. Fig.~\ref{FIG:3}(d) shows the GB structure at 90.4~ns, where a kite phase nucleates transiently between the SK and FK phases. During the simulation, one to four structural units of the kite phase are transiently observed before the two junctions recombined. The kite phase is expected to have $[n] \approx 0$, and the analysis reveals a plateau in the GB atomic excess approaching this value, as shown in Fig.~\ref{FIG:3}(e).

\subsection{Same GB Phases Forming Different GB Phase Junctions}
In this section, we demonstrate that even when a microstructure is composed of only two GB phases, it can still contain different GBPJs with distinct Burgers vectors. Identifying the distinct phases is therefore not sufficient to fully quantify the GB phase microstructure. For example, a GB phase formed by inserting a fraction $[n]$ of a bulk atomic plane can equivalently be created by introducing $1-[n]$ vacancies. Although both routes produce the same GB phase, the GBPJs formed between the parent and nucleated phases differ, which in turn can influence their behavior during migration, deformation, creep and interaction with point defects.

\begin{figure*}[h!]
    \centering
    \includegraphics[width=1.0\textwidth]{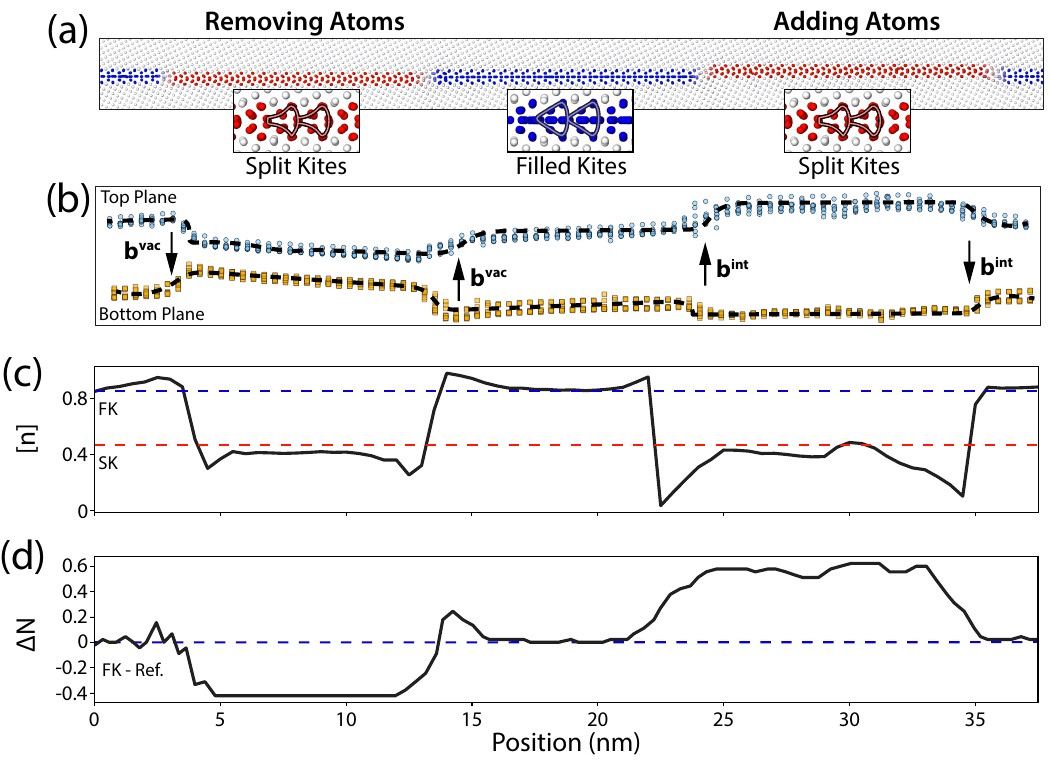}
    \caption{Demonstrating vacancy- and interstitial-type GB phase junctions in a Cu $\Sigma5$(310)[001] tilt boundary. (a) Heterogeneous GB microstructure with FK (blue) and SK (red) phases. The left SK region is nucleated by atom removal, the right by atom insertion. (b) Magnified bulk atomic plane displacement in the top and bottom grains, revealing reduced boundary thickness in the vacancy-induced SK region and increased thickness in the interstitial-induced SK region, with the opposite-sign Burgers vector components $b_3^{\mathrm{vac}}$ and $b_3^{\mathrm{int}}$ indicated. (c,d) Profiles of GB excess number of atoms $[n]$ and the relative change in the number of atoms $\Delta N$. $[n]$ identifies both SK regions as the same GB phase, while $\Delta N$ reveals their distinct formation history.}
    \label{FIG:2}
\end{figure*}

To illustrate this, we consider the Cu $\Sigma5$(310)[001] tilt boundary shown in Fig.~\ref{FIG:2}(a), initialized in a uniform FK parent phase with $[n] = 6/7$. To construct a heterogeneous boundary containing both FK and SK phases separated by distinct GBPJs, we select two $10~\mathrm{nm}$-wide regions along the boundary plane separated by a $10~\mathrm{nm}$-wide FK segment. In the left region, atoms equivalent to $2/5$ of a bulk atomic plane are removed above the GB; in the right 
region, interstitials corresponding to $3/5$ of a bulk atomic plane are inserted. This construction is analogous to introducing vacancy and interstitial loops in a bulk crystal. The system is subsequently equilibrated at $300~\mathrm{K}$ for $5~\mathrm{ns}$, during which both the atom-depleted and atom-enriched regions nucleate the SK phase.

The distinct formation mechanisms of the two SK regions are reflected in the displacements of the bulk atomic planes adjacent to the GB, shown in Fig.~\ref{FIG:2}(b). Atom removal causes the neighboring planes to move closer together, whereas interstitial insertion displaces them farther apart relative to the FK structure. These displacements allow us to quantify the Burgers vector of each GBPJ, yielding $\mathbf{b}^{\mathrm{vac}} = (0.59\,\hat{\mathbf{e}}_1 - 0.45\,\hat{\mathbf{e}}_3)\,\text{\AA}$ and $\mathbf{b}^{\mathrm{int}} = 0.35\,\hat{\mathbf{e}}_3\,\text{\AA}$. The opposite signs and different magnitudes reflect the distinct defect character of the two junctions: one corresponds to a vacancy dipole and the other to an interstitial dipole.

To quantify the excess atomic density of the boundary, we compute both \([n]\) and \(\Delta N\), shown in Fig.~\ref{FIG:2}(c,d). Our analysis shows that both SK regions have \([n]\approx 7/15\), consistent with the expected value (red), while the FK region has \([n]\approx 6/7\), also in agreement with its expected value (blue). Thus, \([n]\) identifies the GB structure, but it does not describe how that structure was formed. To distinguish between SK structures produced by atom removal and insertion, we compute \(\Delta N\) relative to the FK phase, as shown in Fig.~\ref{FIG:2}(d). We find that the left SK region is depleted by \(2/5\) of a bulk atomic plane relative to FK, whereas the right SK region contains an excess of \(3/5\) of a bulk atomic plane. This example shows how these two parameters, related to the density of atoms at the boundary compliment each other and provide a complete description of the microstructure:  \([n]\) identifies the distinct GB phases, while \(\Delta N\) reflects the formation history of each GB phase.

\subsection{GB Phase Junction with a screw component}
\begin{figure*}[h!]
    \centering
    \includegraphics[width=1.0\textwidth]{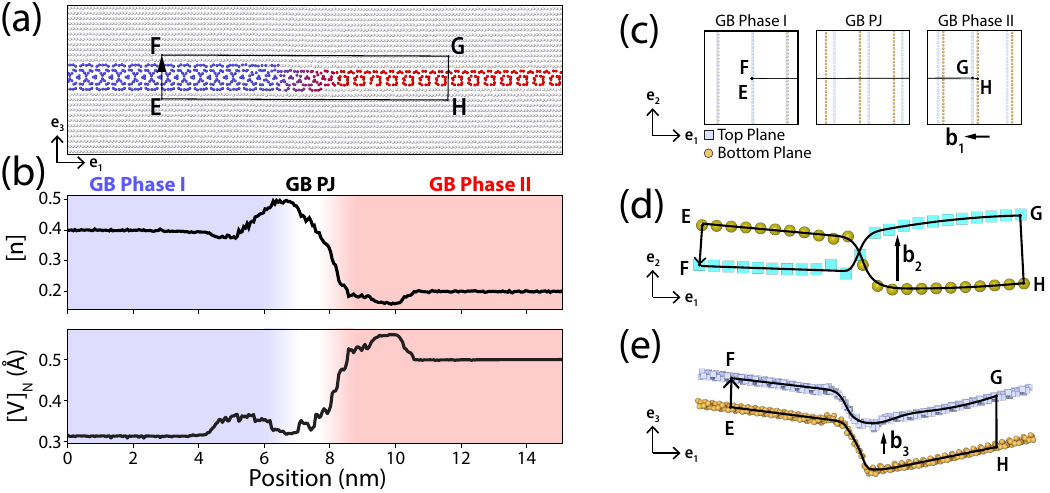}
    \caption{Characterization of a GB phase junction with normal and shear Burgers components in a W $\Sigma5$(310)[001] twist boundary. (a) Atomistic view of two coexisting GB phases separated by a GBPJ, with a Burgers circuit EFGH enclosing the junction. (b) Profiles of GB number of atoms $[n]$, and excess volume $[V]_N$ across the phase junction. (c--e) Scaled atomic displacements of bulk planes in the top and bottom grains viewed along $x$, $y$, and $z$, respectively, revealing the three Burgers vector components $b_1 = -0.15~\text{\AA}$, $b_2 =0.7~\text{\AA}$, and $b_3 = 0.11~\text{\AA}$ of the junction.}
    \label{FIG:4}
\end{figure*}
The GBPJs examined previously, including those in this work, are of edge character, with non-zero Burgers vector components normal and tangential to the boundary plane but no screw component. Consequently, their Burgers vectors can be determined using two-dimensional Burgers circuits. More generally, however, GB microstructures may contain junctions with a nonzero screw component, whose characterization requires a fully three-dimensional construction of the Burgers circuit. Such junctions have not been previously characterized, and this section focuses on their identification and analysis.

Fig.~\ref{FIG:4}(a) shows a W $\Sigma$5(310)[001] twist boundary, modeled using an EAM potential~\cite{zhou2001atomic}, containing a GBPJ, together with the Burgers circuit (EFGH) enclosing the junction. The excess quantities across the phase junction are presented in Fig.~\ref{FIG:4}(b), including  the GB atomic excess $[n]$ and the excess volume $[V]_N$. The profiles indicate that GB Phase I can be obtained from the ground-state structure GB Phase II by inserting $1/5$ of a bulk atomic plane.

Figures~\ref{FIG:4}(c--e) illustrate three different projections of the Burgers circuit constructed around GBPJ in 3D. As before to illustrate the displacement we select two atomic planes above and below the GB. Fig.~\ref{FIG:4}(c) shows the relative positions of atomic rows in the top and bottom planes, viewed normal to the GB plane and scaled along $\mathbf{e}_1$-direction for clarity. The atomic rows are nearly aligned within GB Phase I but are shifted in the $\mathbf{e}_1$-direction within GB Phase II, yielding an in-plane edge component $b_1 = -0.15~\text{\AA}$. This relative displacement appears even larger directly above the junction, likely due to the local distortion due to the GBPJ core.

Fig.~\ref{FIG:4}(d) shows the same atomic rows scaled along the $\mathbf{e}_2$-direction. The upper and lower crystals are shifted relative to each other along the $\mathbf{e}_1$-direction in opposite directions across the two GB phases, yielding a screw component $b_2 = 0.7~\text{\AA}$. Finally, Fig.~\ref{FIG:4}(e) shows the $\mathbf{e}_3$-components of the displacement, revealing a normal Burgers component $b_3 = 0.11~\text{\AA}$ arising from the differences in excess volume and number of GB atoms between the two phases.

\subsection{Two-Dimensional GB Microstructure Containing a Phase Nucleus}

\begin{figure*}[h!]
    \centering
    \includegraphics[width=1.0\textwidth]{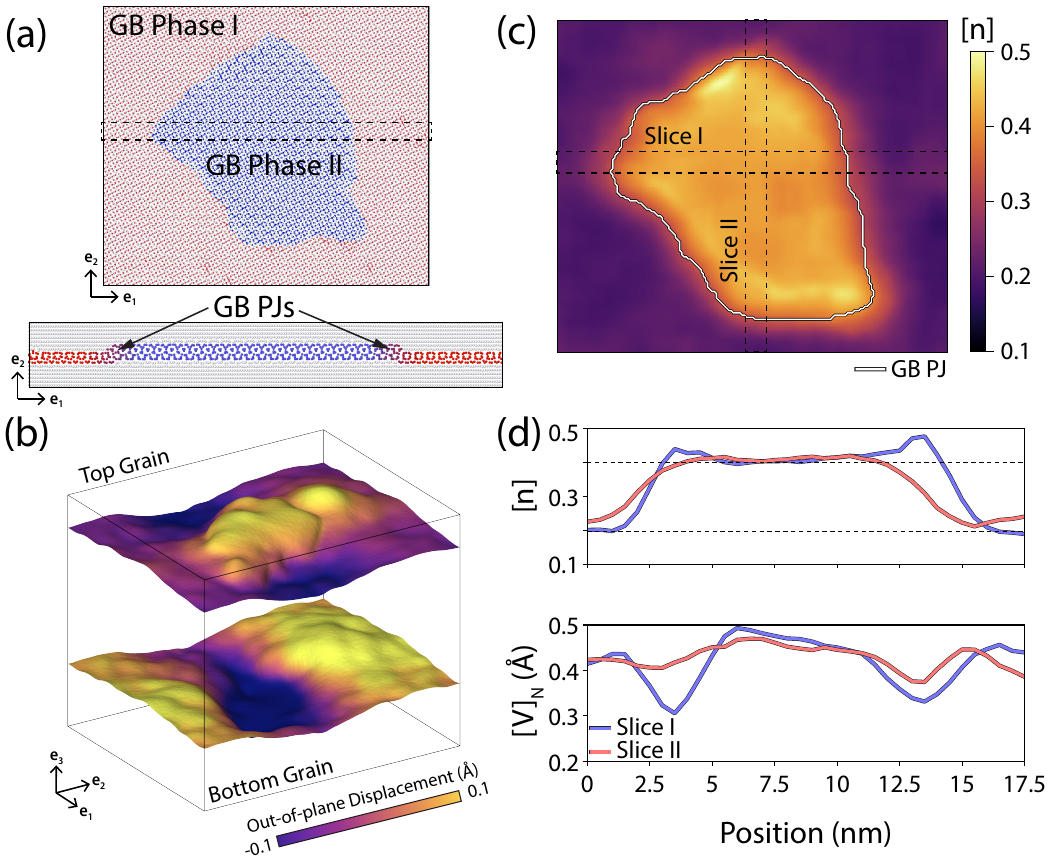}
    \caption{Two-dimensional GB microstructure containing an isolated GB phase nucleus in a W $\Sigma5$(310)[001] twist boundary. (a) Atomistic view of a GB phase nucleus (top) together with a line scan illustrating the distinct structural motifs of the adjoining GB phases. Atoms are colored according to the computed atomic excess, $[n]$, of their corresponding phase, highlighting the different GB phases. (b) Scaled normal displacement of two bulk planes adjacent to the GB. (c) Heat map showing the GB excess of atoms highlighting the embedded phase II and phase junction position (white). (d) Horizontal and vertical line scans of $[n]$ and $[V]_N$ across the nucleus, recovering the expected phase-specific values and their variation across the GBPJ. }
    \label{FIG:5}
\end{figure*}

Building on the analysis of quasi-two-dimensional systems in the preceding sections, where GBPJs extend across the periodic simulation cell, we now analyze a fully three-dimensional GB microstructure of the same W $\Sigma5$(310)[001] twist boundary, in which a localized nucleus of GB 
Phase II is embedded within the parent GB Phase I, as shown in Fig.~\ref{FIG:5}(a). The heterogeneous structure of this boundary was previously simulated in Ref.~\cite{frolov_2018}, but the nucleus was not visualized or characterized. The structure is generated by introducing interstitial atoms above the ground-state GB Phase I, which, during subsequent annealing for $\approx 200~\mathrm{ns}$, diffuse into the boundary and trigger a first-order GB phase transition. The nucleus exhibits a non-circular morphology with well-defined facets, clearly visible in Fig.~\ref{FIG:5}(a), likely reflecting anisotropy of the GBPJ line energy. This interfacial microstructure is analogous to interstitial loop formation in bulk crystals, with the key distinction that the bounding line defect is a GBPJ carrying a smaller Burgers vector than a perfect disconnection and enclosing a region of distinct GB structure.

Both interstitial- and vacancy-type GB loops could produce the morphology shown in Fig.~\ref{FIG:5}(a), since a nucleus of the same GB phase could, in principle, also be generated by introducing vacancies. The algorithm identifies the loop type by locating the nucleus, quantifying the excess properties of the two GB structures, and measuring differences in local atomic density. Fig.~\ref{FIG:5}(b) shows bulk atomic planes taken above and below the GB, colored by their out-of-plane ($\mathbf{e}_3$) displacement; the distortion of these planes reveals that the GB phase constituting the nucleus has a larger thickness. 
Next, we compute the excess number of GB atoms, $[n]$, as a function of position on the boundary plane, shown as a heat map in Fig.~\ref{FIG:5}(c). The map clearly identifies the nucleus and reveals the higher $[n]$ associated with the structure of GB Phase II. The transition in local atomic density between the values of GB Phase I and Phase II identifies the position of the GBPJ, shown as a white line in Fig.~\ref{FIG:5}(c). The sharp faceted GBPJ segments apparent in the atomistic view in Fig.~\ref{FIG:5}(a) appear somewhat smoothed in the heat map due to spatial averaging. The heat map of the relative change in atomic number, $\Delta N$, is qualitatively similar to that of $[n]$ and is therefore not shown, indicating a higher density of atoms within the nucleus and confirming the interstitial character of the GB loop. Finally, line scans taken along horizontal (blue) and vertical (red) directions, shown in Fig.~\ref{FIG:5}(d), of the GB atomic excess (top) and excess volume (bottom) match the expected values within each phase and illustrate their variation across the GBPJ.

\section{Discussion}

Grain boundaries have traditionally been treated as structurally uniform interfaces, described by a single structure for a given misorientation and boundary plane~\cite{sutton1995interfaces}. Because resolving the interior of a boundary which may contain several coexisting GB phases remains experimentally difficult, little is known about the conditions under which such microstructures should be expected. As a result, the concept of a GB phase microstructure has emerged only recently, following the demonstration of first-order GB phase transitions and equilibrium coexistence by atomistic simulations~\cite{frolov2013structural,zhu2018predicting,frolov2018structures} and by direct atomic-resolution imaging~\cite{merkle1987atomic,peter2018segregation,GBPhaseExp,langenohl2022dual,devulapalli2024topological}, mostly in quasi-2D geometries. While it is not clear whether multiphase GB microstructures should be viewed as rare special cases or as generic features of evolving GBs, heterogeneous boundaries can form near GB phase coexistence, but they should arise far more frequently during non-equilibrium processes.

First, two-phase and multiphase states should be expected near first-order transitions, when the competing phases are close to equilibrium and have nearly equal free energies. Their existence, however, is expected only within a relatively narrow thermodynamic interval. For example, for a congruent transition in an elemental system at fixed stress, Gibbs phase rule restricts coexistence to a single temperature, while in multicomponent systems the coexistence region spans a finite but still limited range of temperatures and compositions~\cite{frolov2015phases}. Equilibrium coexistence can therefore explain only a subset of GB microstructures, namely those formed near GB phase boundaries in the interfacial phase diagram~\cite{luo2023computing}.

That such states are physically realized is nevertheless supported by both theory and experiment. Elasticity theory predicts that alternating GB phase patterns can be stabilized by long-range elastic interactions between the line defects and line forces associated with the GBPJs~\cite{winter2024phase}. In atomistic simulations, these interactions were shown to stabilize periodic patterns of alternating GB phases over microsecond time scales without an external driving force~\cite{winter2024phase}. Experimentally, alternating domino and pearl segments were observed along a Cu $\Sigma37 \langle111\rangle$ boundary by STEM~\cite{langenohl2022dual}. This latter observation was made below the calculated transformation temperature and its stability was attributed to the limited mobility of the junctions potentially combined with their elastic interaction with pre-existing disconnections~\cite{langenohl2022dual}. Nevertheless, these studies show that patterned GB microstructures can form and persist under near-equilibrium or weakly driven
conditions.

Interfacial microstructures should be considerably more common under non-equilibrium conditions,
when the boundary participates actively in a materials process~\cite{wei2021direct}. During grain growth at high temperature, mechanical deformation by GB sliding and shear-coupled migration~\cite{cahn2004unified,khater2012disconnection}, diffusional creep sustained by climb of GB dislocations, or irradiation, which exposes boundaries to fluxes of point defects and can nucleate GB prismatic loops~\cite{frolov2018structures,han2015interplay}, the boundary changes its position, shape, and atomic content. Such changes are conventionally assumed to be accommodated by the nucleation and motion of disconnections, whose Burgers vectors belong to the DSC lattice and which preserve a single GB structure~\cite{chen2020temperature,khater2012disconnection,thomas2026grain,han2018grain,gautier2025quantifying}.

It has recently been shown, however, that the Burgers vector of a GBPJ is confined to the Wigner--Seitz cell of the DSC lattice and therefore can always be smaller in magnitude than that of a
regular disconnection~\cite{winter2025quantifying}. Because the elastic energy of a line defect scales with the square of its Burgers vector, GBPJs can therefore be energetically competitive with, or even favored over, conventional disconnections, even though their formation necessarily creates a patch of a second GB phase. For example, for non-climb processes such as pure sliding, the coexistence condition between two GB phases $\alpha$ and $\beta$ must include the work of the Peach--Koehler force acting on the DSC component of the defect separating them~\cite{winter2025quantifying}:
\begin{equation}
    \gamma^{\alpha} - \gamma^{\beta} = \sum_{i=1}^{2}\sigma_{i3}\,b_{i}^{DSC},
    \label{eq:pk}
\end{equation}
where $\sigma_{i3}$ are the resolved shear stresses on the boundary plane and $b_i$ are the in-plane DSC components of the Burgers vector of the separating junction. This relation shows that the driving force for a GB phase transformation is not determined by the free-energy difference alone, but by the combined effect of $\Delta\gamma$ and the Peach--Koehler work. Because many competing GB phases are nearly degenerate in energy, modest applied stresses can offset the free-energy penalty of a metastable phase and promote its formation through GBPJ nucleation and motion. Under these conditions, multiphase GB microstructures should emerge even away from equilibrium coexistence. In this respect GBPJs play a role analogous to partial dislocations in bulk crystals.

It should therefore be expected that non-equilibrium processing generates complex interfacial microstructures composed of distinct GB phases separated by networks of GBPJs. Complex multiphase GB microstructures should therefore be regarded as the expected state of boundaries in a material under processing or service conditions, not as an exotic near-coexistence
phenomenon.

The framework introduced in this work automates the identification and analysis of GB phase microstructures. Distinct GB phases are identified from spatially resolved excess properties computed with a moving analysis window, the junctions between them are located from the distortion of the bulk planes adjacent to the boundary, and the Burgers content of each junction is estimated from Burgers circuits constructed in the current configuration, so that no relaxed reference structures are required.

Two limitations of the present analysis should be noted. First, the Burgers vectors obtained here are approximate measures of the lattice mismatch across a junction, because they are evaluated from circuits constructed in the current configuration rather than from an exact topological reference state. Second, the present implementation is restricted to planar CSL boundaries, and extension to more general curved or faceted interfaces will require additional development. Within these limits, however, the method provides the quantities needed to analyze the nucleation, growth, coarsening, and possible spinodal decomposition of GB phases, and it creates a practical route toward measuring GBPJ mobilities and linking GB phase behavior to migration, sliding, creep, and radiation damage.

\section*{Supplemental Material}
The Supplemental Material includes a figure illustrating the sensitivity of the calculated Burgers vector of a grain boundary phase junction to the analysis window size, as well as a video showing the split-kite–to–filled-kite phase transition in Cu.

\section*{Acknowledgments}
This work was performed under the auspices of the U.S. Department of Energy (DOE) by Lawrence Livermore National Laboratory under contract DE-AC52-07NA27344. TF was supported fully and DM partially by the U.S. DOE, Office of Science under an Office of Fusion Energy Sciences Early Career Award. RER was supported by the U.S. DOE, Office of Science, Office of Fusion Energy Sciences. ISW received support from the LDRD program at Sandia National Laboratories. Computing support for this work came from the Lawrence Livermore National Laboratory Institutional Computing Grand Challenge program. An award of computer time was also provided by the INCITE program. This research used resources of both the Argonne and Oak Ridge Leadership Computing Facilities, which are DOE Office of Science User Facilities supported under contracts DE-AC02-06CH11357 and DE-AC0500OR22725, respectively. This article has been authored by an employee of National Technology \& Engineering Solutions of Sandia, LLC under Contract No. DE-NA0003525 with the U.S. Department of Energy (DOE). The employee owns all right, title and interest in and to the article and is solely responsible for its contents. The United States Government retains and the publisher, by accepting the article for publication, acknowledges that the United States Government retains a non-exclusive, paid-up, irrevocable, world-wide license to publish or reproduce the published form of this article or allow others to do so, for United States Government purposes. The DOE will provide public access to these results of federally sponsored research in accordance with the DOE Public Access Plan https://www.energy.gov/downloads/doe-public-access-plan. This paper describes objective technical results and analysis. Any subjective views or opinions that might be expressed in the paper do not necessarily represent the views of the U.S. Department of Energy or the United States Government.




\bibliographystyle{unsrt} 
\bibliography{cas-refs}





\end{document}